\documentclass[aps,twocolumn,pra,tightenlines,floatfix,showpacs,superscriptaddress,longbibliography]{revtex4-1}
\usepackage{graphicx}
\usepackage{epstopdf}
\usepackage[english]{babel}
\usepackage{amsmath}
\usepackage{amssymb}
\usepackage{mathrsfs}
\usepackage[none]{hyphenat}
\usepackage{appendix}
\usepackage{bm}
\usepackage{float}
\usepackage{color}
\usepackage{longtable}
\usepackage{url}
\usepackage[usenames,dvipsnames]{xcolor}
\usepackage[colorlinks=true,linkcolor=Blue,urlcolor=Blue,citecolor=Blue]{hyperref}

\graphicspath{{./Figures}}

\begin{document}

\title{Estimating the Euclidean quantum propagator with deep generative modeling of Feynman paths}

\author{Yanming Che}
\email{yanmingche01@gmail.com}
\affiliation{Department of Physics, University of Michigan, Ann Arbor, Michigan
48109-1040, USA}
\affiliation{Theoretical Quantum Physics Laboratory, RIKEN Cluster for Pioneering
Research, Wako-shi, Saitama 351-0198, Japan}

\author{Clemens Gneiting}
\email{clemens.gneiting@riken.jp}
\affiliation{Theoretical Quantum Physics Laboratory, RIKEN Cluster for Pioneering
Research, Wako-shi, Saitama 351-0198, Japan}
\affiliation{RIKEN Center for Quantum Computing (RQC), 2-1 Hirosawa, Wako-shi, Saitama 351-0198, Japan}

\author{Franco Nori}
\email{fnori@riken.jp}
\affiliation{RIKEN Center for Quantum Computing (RQC), 2-1 Hirosawa, Wako-shi, Saitama 351-0198, Japan}
\affiliation{Theoretical Quantum Physics Laboratory, RIKEN Cluster for Pioneering
Research, Wako-shi, Saitama 351-0198, Japan}
\affiliation{Department of Physics, The University of Michigan, Ann Arbor, Michigan
48109-1040, USA}

\date{\today}

\begin{abstract}
Feynman path integrals provide an elegant, classically-inspired 
representation for the quantum propagator and the quantum dynamics, through 
summing over a huge manifold of all possible paths. 
From computational and simulational perspectives, the ergodic tracking 
of the whole path manifold is a hard problem. Machine learning can help, in 
an efficient manner, to identify the relevant subspace and the intrinsic 
structure residing at a small fraction of the vast path manifold. 
In this work, we propose the Feynman path generator for quantum mechanical systems, 
which efficiently generates Feynman paths with fixed endpoints, from a 
(low-dimensional) latent space and by targeting a desired density of paths in the 
Euclidean space-time. With such path generators, the Euclidean propagator as well 
as the ground state wave function can be estimated efficiently for a generic potential energy. 
Our work provides an alternative approach for calculating the quantum propagator and the ground state wave function, paves the way toward generative modeling of quantum mechanical Feynman paths, and offers a different perspective to understand the quantum-classical correspondence through deep learning.

\end{abstract}


\maketitle
\section{Introduction}
Feynman path (FP) integrals provide an elegant, classically-inspired representation 
for the quantum propagator and the quantum dynamics, through 
summing over a huge manifold of all possible paths connecting two fixed endpoints~\cite{Feynman_RMP_1948,RichardArxiv2000,Feynman_Hibbs_book}. From the perspective of the path integral, 
quantum dynamics arises from coherent contributions including both the classical path and quantum fluctuations. It thus also provides an intuitive framework for understanding the quantum-classical correspondence. 

Generally, the ergodic tracking of each path contribution to the quantum propagator is a computationally hard problem. In this respect, a similar situation occurs in the classical simulation of many-body quantum systems, whose Hilbert space is exponentially large, prohibiting exact and efficient methods in the generic case. However, in many practical scenarios, physically relevant features and structures may only reside at a small fraction of the vast Hilbert space; and this enables efficient machine learning techniques for quantum-physics problems, like the neural-network approach for quantum many-body systems~\cite{CarleoTroyerScience2017,Carrasquilla2017,GaoNC2017,EquivPRB2017,NieuwenburgNatPhys2017,Schuld2014,Dunjko2018,CarleoRMPML,LuchnikovEntropy2019,YoshiokaPRB2019,VicentiniPRL2019,NagyPRL2019,HartmannPRL2019,Berezutskii2020,TanPRA2021}, neural network quantum state tomography~\cite{TorlaiNatPhys2018,MelkaniPRA2020,Palmieri_2020,LohaniMLST2020,NeugebauerPRA2020,ShahnawazArxiv2020,*ShahnawazPRR2021}, manifold learning and clustering of quantum phases~\cite{ChePRB2020,ScheurerPRL2020,LongPRL2020,Mendes-SantosPRX2021,Zohar2012,arxiv2106.12627}, and physical concepts rediscovered with neural networks~\cite{ItenPRL2020,WangScienceBulletin2019}.

Similarly, in the path integral formalism, contributions from large deviations with respect to the classical path generically cancel each other due to rapid oscillations (in the Lorentzian space-time), or are exponentially suppressed (in the Euclidean space-time)~\cite{Feynman_RMP_1948,Feynman_Hibbs_book,RichardArxiv2000}. Therefore, dominant contributions to the propagator may only reside at a corner of the huge path manifold.
Then one may naturally ask: $(1)$ Can machine learning capture these structures in FPs? $(2)$ Can the propagator be efficiently estimated with neural networks? So far, the literature relating machine learning to FP integrals either has not fully investigated these questions~\cite{LiuPRL2020,arxiv:1904.07568,Barr2020,arxiv:2011.11891,Mendes-SantosPRXQ2021}, or has mainly focused on quantum field-theoretical systems~\cite{PhysRevLett.121.260601,AlbergoPRD2019,PhysRevResearch.2.023369,PhysRevLett.125.121601,NicoliPRL2021,PhysRevD.104.114507,MatijaArxiv2020,arxiv:2201.08862v2}. In this paper, we address these questions by proposing the FP generator for quantum mechanical systems, which provides a fresh perspective on efficiently calculating the propagator as well as the ground state wave function, and on studying FPs with deep generative modeling.
\begin{figure}[t]
\centerline{\includegraphics[height=1.6in,width=3.2in,clip]{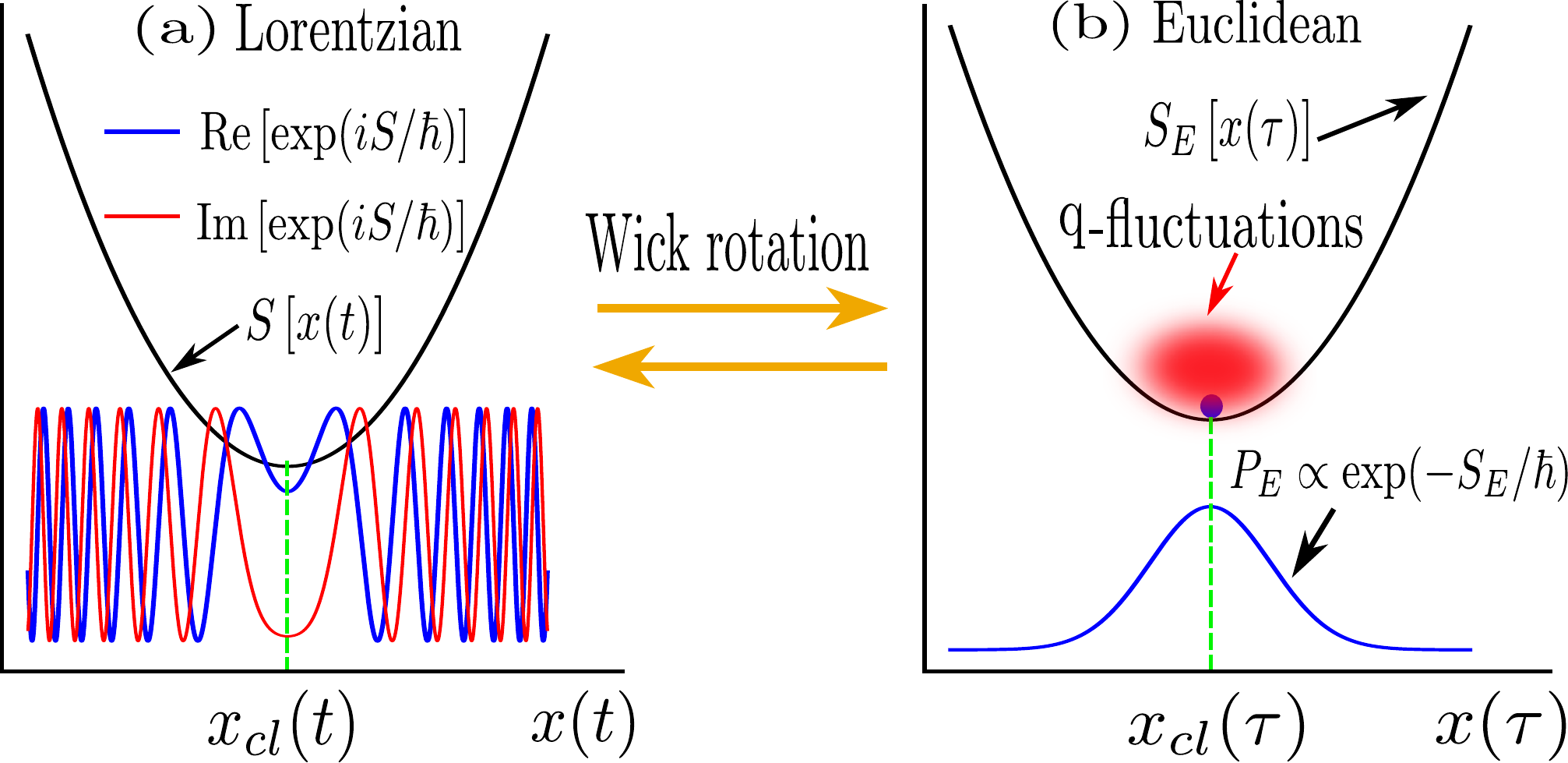}}
\caption{\textbf{Schematics of Feynman paths and path integrals in (a) Lorentzian and (b) Euclidean space-times.} 
The times $\tau$ and $t$ are connected via the Wick rotation $it \rightarrow \tau$. The integrands of the propagators are plotted with the respective representative action curves $S\left[ x(t)\right]$ and $S_E\left[ x(\tau)\right]$. In (a), $\mathrm{Re}$ and $\mathrm{Im}$ denote the real and imaginary parts, respectively, and $x_{cl} (t)$ stands for the classical path (stationary path), which satisfies $\delta S = 0$~\cite{Feynman_Hibbs_book}. In (b), $x_{cl} (\tau)$ is the path with the least Euclidean action. It can be seen that such paths and their neighbourhoods contribute dominantly to the propagators, while large deviations away from them cancel each other through rapid oscillations in (a), and are exponentially suppressed in (b). In (b), the exponential decay factor defines a probability distribution function for the Feynman paths in Euclidean space-time. In the case when $x_{cl} (\tau)$ is identical to the classical Euclidean path, the red cloud around it represents quantum fluctuations. 
}
\label{fig:FPI_Lorentzian_Euclidean}
\end{figure}

\section{Lorentzian and Euclidean Feynman paths} 
We start from a brief overview of the real- and imaginary-time FP integral. The quantum unitary time evolution of the wave function is given by 
\begin{equation}
\psi \left(x_f, t_f \right) 
= \int \mathrm{d}x_i \; K \! \left(x_f, t_f; x_i, t_i \right) \psi \! \left(x_i, t_i \right),
\end{equation}
where $(x, t)$ denotes the $\left(d+1\right)$ dimensional space-time, and $\psi(x_i, t_i)$ is the initial wave function. 
In the FP integral formulation, the propagator (or kernel) is given by summing over all possible paths starting from $x_i$ at time $t_i=0$ and ending at $x_f$ at time $t_f$~\cite{Feynman_RMP_1948,Feynman_Hibbs_book,RichardArxiv2000,Lin_FN_PRB2002,*Lin_FN_PRL1996,*Lin_FN_PRB1994}:
\begin{equation} 
K \left(x_f, t_f; x_i, 0 \right) = A(t_f) \sum_{\{x(t)\}} \mathrm{exp} \left( i S\left[ x(t)\right] 
/ \hbar \right),
\label{eq:Propagator_Lorentzian}
\end{equation}
where $A(t_f)$ is the normalization factor independent of the specific path; ${\{x(t)\}}$ is the manifold of all possible FPs connecting $\left(x_i, 0 \right)$ and $\left(x_f, t_f\right)$ in space-time; $S\left[x(t)\right]$ is the classical action of the path $x(t)$; and $\hbar$ is the Planck constant.

An equivalent but more convenient formalism for numerical evaluation is to work in the Euclidean space-time, via the Wick rotation $it \rightarrow \tau$. Then the system propagates along the imaginary time $\tau$, and the propagator at $\tau = T$ reads
\begin{equation}
\label{eq:Propagator_Euclidean}
K_E \left(x_f, T; x_i, 0 \right) = A_E(T) \sum_{\{x(\tau)\}} \mathrm{exp} \left( -S_E\left[x(\tau)\right] / 
\hbar \right),
\end{equation}
where $x(\tau)$ is the Euclidean FP and 
\begin{equation}
S_E \left[x(\tau)\right] = \int_0^T H_E \left[x(\tau)\right] \mathrm{d}\tau
\end{equation}
is the Euclidean action, with 
\begin{equation}
H_E \left[x(\tau)\right] = \frac{1}{2}m \left( \frac{\mathrm{d}x}{\mathrm{d}\tau}\right)^2 + V \left[x(\tau)\right] 
\end{equation}
the Hamiltonian and $V (x)$ the potential energy.

From the Lorentzian and Euclidean path integrals, one can find sparse features and patterns in the path manifold (Fig.~\ref{fig:FPI_Lorentzian_Euclidean}). The dominant contributions to the propagator come from paths that are located at the basin around the stationary path (which satisfies $\delta S = 0$) in the action landscape~\cite{stationary_note}. 
The stationary path together with quantum fluctuations in the vicinity of it dictate the quantum kernel, while large deviations cancel each other due to rapid oscillations in the real-time path integral, and are exponentially suppressed in the Euclidean version (\ref{eq:Propagator_Euclidean}).

Equation (\ref{eq:Propagator_Euclidean}) can be interpreted as a probability distribution of FPs in Euclidean space-time,
\begin{equation}
\label{eq:PDF_Euclidian_FP}
P_E \left[ x(\tau) \right] = Z^{-1} \mathrm{exp} \left( -S_E\left[x(\tau)\right] / \hbar \right),
\end{equation}
where 
\begin{equation}
Z = \sum_{\{x(\tau)\}} \mathrm{exp} \left( -S_E\left[x(\tau)\right] / \hbar \right)
\end{equation}
is the partition function. Note that here $Z$ depends on the two endpoints. When the normalization factor $A_E(T) =1$, $Z$ is the imaginary-time propagator in Eq.~(\ref{eq:Propagator_Euclidean}). Here $P_E \left[ x(\tau) \right]$ is the target distribution of our FP generator.

\begin{figure}[b]
\centerline{\includegraphics[height=2.2in,width=3in,clip]{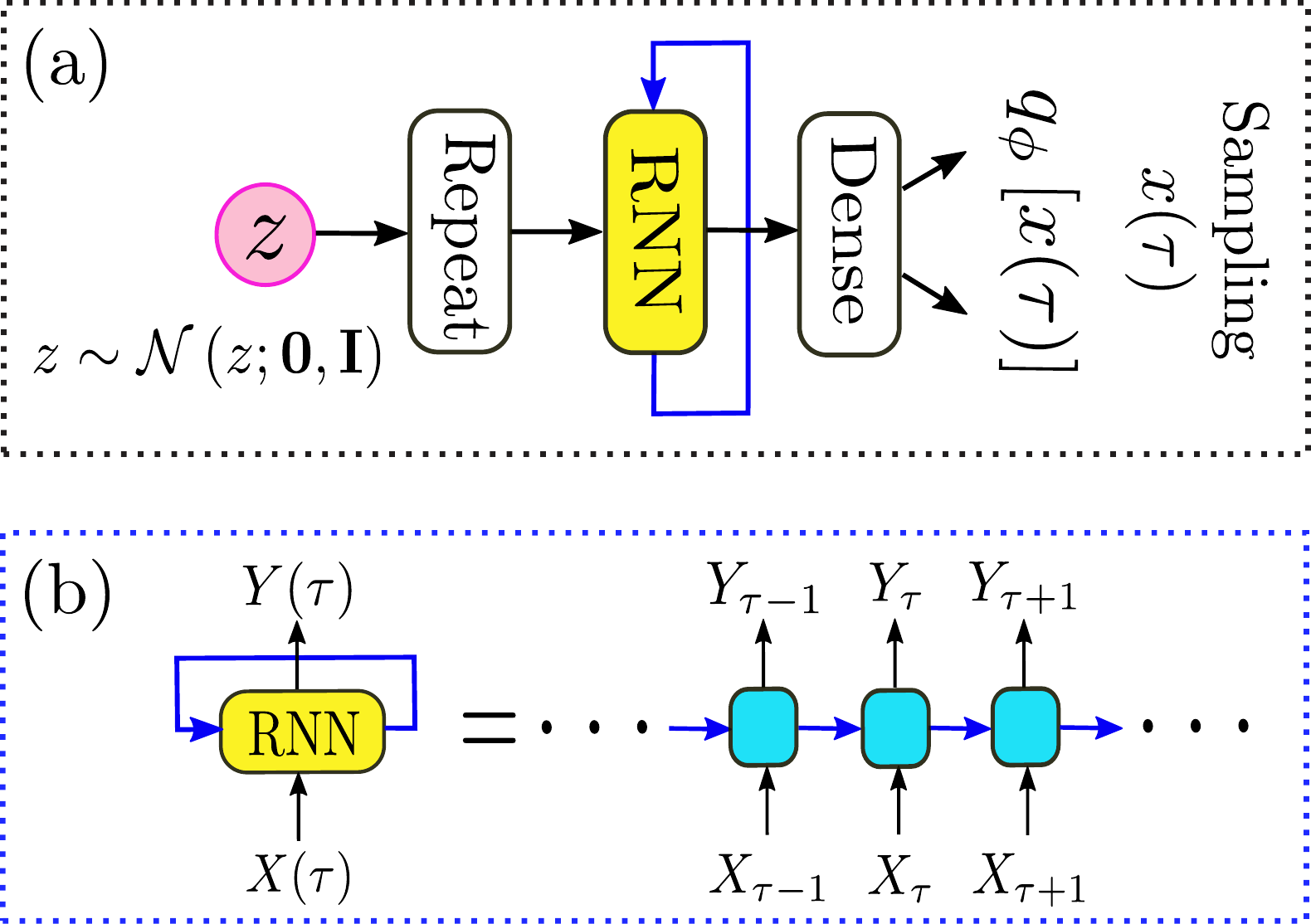}}
\caption{\textbf{(a) A variational realization of the Feynman path generator (VFPG) with a recurrent neural network (RNN).} 
The input of the VFPG is a $2\mathrm{D}$ latent vector $z$ sampled from the standard Gaussian distribution ${\cal{N}} \left(z; \mathbf{0}, \mathbf{I} \right)$, followed by a repeat operation (Repeat) that extends the dimension to match the input shape of the RNN. The output sequence is fed into densely connected layers (Dense) to produce the variational parameters $\phi$ for the path distribution $q_{\phi} \left[x(\tau)\right]$, which is modeled, at each time stamp, by a Gaussian mixture (see Appendix~\ref{sec:A} for details). The network is trained to target the path distribution in Eq.~(\ref{eq:PDF_Euclidian_FP}). 
\textbf{(b)} The structure of the RNN in (a), where $X(\tau)$ and $Y(\tau)$ are the input and output sequences, respectively; the blue squares on the right denote the RNN units; and the blue arrows denote the hidden state transfer~\cite{GoodfellowDL2016}. In this work we use the long-short term memory (LSTM)~\cite{LSTM} for the RNN.
}
\label{fig:VFPG_arch}
\end{figure}

\section{Variational Feynman path generator} 
We now show that the Euclidean quantum propagator can be efficiently estimated through the generative modeling of FPs. We first introduce the concept of the FP generator, 
which produces FPs with fixed endpoints from a latent space, in which tracking and sampling are made simple. The statistics of the latent variables can be modeled by simple distributions such as the standard Gaussian. Such FP generators can be realized through variational recurrent neural networks (RNN)~\cite{PixelRNN2016} or through normalizing flows~\cite{NICE2014,RealNVP2016}.

In this work, we will restrict ourselves to the Euclidean space-time, where the density of paths is described by the target distribution (\ref{eq:PDF_Euclidian_FP}), and will use a variational FP generator (VFPG) of deep neural-net architecture, to approximate the target distribution and generate FPs via an efficient parallel sampling.
As shown in Fig.~\ref{fig:VFPG_arch}, the architecture of the VFPG is similar to that of the decoder in the variational autoencoder (VAE), a well-known generative model for learning image and time-series signals~\cite{VAE2013,DoerschArxiv2016Tutorial,ChungMethods2019,ChungNeurIPS2015}.

Here a FP $x(\tau)$ is represented by a discrete $N_{\tau}$-dimensional sequence (trajectory) $\{ x \left( \tau_k \right) \}^{N_{\tau} - 1 }_{k=0}$, with $\tau_k = k \times \Delta \tau$, where $\Delta \tau = T/\left( N_{\tau} - 1 \right)$ is the time step (see Appendix~\ref{sec:A} for details).
The VFPG samples $2\mathrm{D}$ latent vectors $z$ from the standard Gaussian distribution ${\cal{N}} \left(z; \mathbf{0}, \mathbf{I} \right)$ as its input and generates the path time series $x(\tau)$ with fixed endpoints $(x_i, 0)$ and $(x_f, T)$, respectively, according to the output variational density distribution of paths $q_{\phi}\left[ x(\tau) \right]$, with $\phi$ the output of the VFPG. The most probable output path should be the one that minimizes $S_E$.

As pictured in Fig.~\ref{fig:VFPG_arch}, the output sequence of the RNN is fed into a densely connected layer to produce the variational parameters $\phi$, and at each time stamp, $q_{\phi} \left[x(\tau)\right]$ is modeled as a Gaussian mixture model, which is a universal approximator of densities~\cite{GoodfellowDL2016} (see Appendix~\ref{sec:A}).
We use the long-short term memory (LSTM)~\cite{LSTM} for the RNN to build the generative network.

\section{Loss Function and Training}
The loss function of the network is given by the Kullback–Leibler (KL) divergence between $q_{\phi}[x(\tau)]$ and $P_E \left[ x(\tau) \right]$ in Eq.~(\ref{eq:PDF_Euclidian_FP}):
\begin{eqnarray}
\label{eq:Loss}
{\cal{L}}_{\mathrm{VFPG}} = D_\mathrm{KL} \left\{ q_{\phi}[x(\tau)] \ || \ P_E \left[ x(\tau) \right] \right\}, 
\end{eqnarray}
subject to the constraints: 
\begin{equation}
x(0) = x_i, \ \ x(T) = x_f.
\end{equation}
The definition of the KL divergence is 
\begin{equation}
D_\mathrm{KL} \left[ q(x) \ || \ p(x) \right] = \mathop{\mathbb{E}}_{x \sim q(x)} \ln \frac{q(x)}{p(x)},
\end{equation}
where $\mathop{\mathbb{E}}$ denotes the expectation value. 

After some straightforward derivations with the definition of the KL divergence and the expression of the target probability $P_E \left[ x(\tau) \right]$ in Eq.~(\ref{eq:PDF_Euclidian_FP}), the VFPG loss can be rewritten as 
\begin{eqnarray}
\label{eq:Generator_Loss}
{\cal{L}}_{\mathrm{VFPG}} = \hbar^{-1} \left\{ \mathop{\mathbb{E}}_{x \sim q_{\phi}} S_E \left[ x(\tau) \right]
+ \hbar \mathop{\mathbb{E}}_{x \sim q_{\phi}} \ln q_{\phi}[x(\tau)] \right\} + \ln Z. \nonumber \\
\end{eqnarray}
The last term in Eq.~(\ref{eq:Generator_Loss}) is independent of the network parameters and does not contribute to the gradient of the loss. The second expectation term is the (negative) differential entropy of the variational distribution $q_{\phi}$. 
\begin{figure*}[t]
\centerline{\includegraphics[height=5in,width=6.5in,clip]{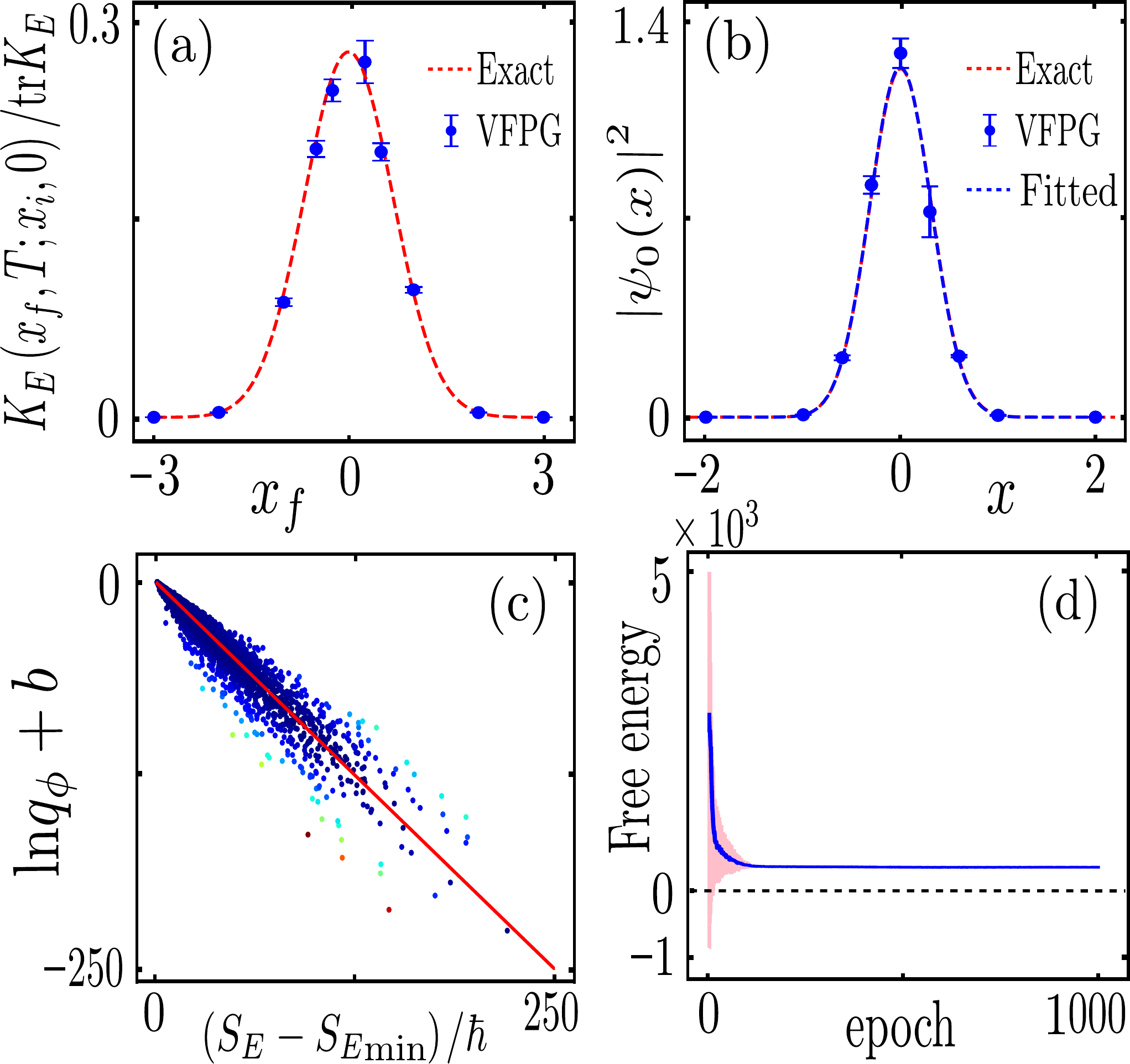}}
\caption{\textbf{Estimating the Euclidean quantum propagator through the variational Feynman path generator (VFPG) for a harmonic oscillator}, where $V(x) = \frac{1}{2} m \omega^2 x^2$.
\textbf{(a)} The trace-normalized propagator vs $x_f$, with $x_i = 0$, $\omega = 1$ and $T = 0.5$.
\textbf{(b)} The ground-state probability density obtained from the Euclidean propagator. The blue dashed line is fit from the VFPG observations (blue dots), and we have used $\omega = 5$ and $T = 1$. In (a, b), the error bars denote the two-standard-deviation of the plotted quantity, evaluated from the training noise (standard deviation) of ${\cal{F}}_{\phi}$. See the main text.
\textbf{(c)} An example of the distribution of $10^4$ generated Feynman paths in the plane of the variational log-probability and the action $S_E$, where both axes are shifted, with $S_{E \mathrm{min}}$ the minimum action and $b = \left(- {\cal{F}}_{\phi} + S_{E \mathrm{min}} \right)/\hbar$. The red solid line represents the exact exponential distribution (\ref{eq:PDF_Euclidian_FP}). It is also the location of Feynman paths satisfying $F_{\phi} \left[x(\tau)\right] = {\cal{F}}_{\phi}$, and the color (from blue to red) indicates the distance between $F_{\phi} \left[x(\tau)\right] $ and its expectation value ${\cal{F}}_{\phi}$. 
\textbf{(d)} An example of the free energy ${\cal{F}}_{\phi}$ (blue line) during the training process for estimating the quantum propagator in (a), where the width of the red shaded area denotes its two-standard-deviation in the generated path ensemble. The parameters used in (c, d) are the same as that in (a), except that the final positions are fixed at $x_f = 1$. Other representative parameters used are: $N_{\tau} = 32$, $\hbar = 1$ and $m = 1$, respectively.
}
\label{fig:HO_Gen}
\end{figure*}

The generator is trained to effectively find an equilibrium for the competition between minimizing the averaged action and maximizing the differential entropy. Viewing the action as an energy functional of a path configuration, and $\hbar$ as the temperature (setting the Boltzmann constant $k_B = 1$), the quantity in the large brackets in Eq.~(\ref{eq:Generator_Loss}) describes the variational free energy of the path manifold (ensemble): 
\begin{equation}
{\cal{F}}_{\phi} = \mathop{\mathbb{E}}_{x \sim q_{\phi}} F_{\phi} \left[x(\tau)\right],
\end{equation}
where 
\begin{equation}
F_{\phi} \left[x(\tau)\right] = S_E \left[x(\tau)\right]+ \hbar \ln q_{\phi}[x(\tau)].
\end{equation}
The true free energy is given by ${\cal{F}} = -\hbar \ln Z$. By taking the normalization in (\ref{eq:Propagator_Euclidean}) to be $A_E (T) = 1$, we can estimate the Euclidean quantum propagator with $K_E \left( x_f, T; x_i, 0\right) = Z_{\phi}$, that is 
\begin{eqnarray}
\label{eq:est_prop}
K_E \left( x_f, T; x_i, 0\right) = \mathrm{exp} \left( - {\cal{F}}_{\phi} / \hbar \right). 
\end{eqnarray}
This means that we used an estimator $Z_{\phi}$ for $Z$ which sets ${\cal{L}}_{\mathrm{VFPG}} = \ln \left( Z/Z_{\phi} \right)$.

Note that the generator loss~(\ref{eq:Generator_Loss}), in terms of the gap between the variational and the true free energies, was also used for addressing the statistical mechanics of lattice spin 
models~\cite{PhysRevLett.121.260601,WuPRL2019,WangArxiv2020}, for sampling molecular structures~\cite{NoeScience2019} and lattice field theories~\cite{PhysRevLett.121.260601,AlbergoPRD2019,PhysRevResearch.2.023369,PhysRevLett.125.121601,NicoliPRL2021,PhysRevD.104.114507,MatijaArxiv2020,arxiv:2201.08862v2}, and for the variational neural annealing~\cite{Hibat-Allah2101}. For the VFPG here, the latent space is low-dimensional and the path manifold is continuous, bringing new challenges for modeling the output density distribution of paths. In addition, the constraints on the two endpoints of each path will be considered as penalties during the training process~\cite{Vadlamani2020} (see Appendix~\ref{sec:B} for details). 

\section{Generating Feynman paths and estimating the quantum propagator} 
Once the VFPG is trained, it can generate FPs by sampling the latent variables $z \sim {\cal{N}} \left(z; \mathbf{0}, \mathbf{I} \right)$, giving an estimation of the Euclidean propagator $K_E \left(x_f, T; x_i, 0 \right)$, which is the kernel of the imaginary-time propagation. In the demonstrated examples, we note that the choice of the number of time stamps $N_{\tau}$ only affects the normalization prefactor of the propagator that is independent of FPs. Thus the normalized propagator will not depends on the value of $N_{\tau}$. 

The spectral representation of the kernel is given by~\cite{InferringDM2019} 
\begin{equation}
K_E \left(x_f, T; x_i, 0 \right) = \sum_{n=0} \mathrm{e}^{-T E_n / \hbar} \psi_n (x_i) \psi_n^{*} (x_f),
\end{equation}
where $E_n$ is the $n$-th eigenenergy and $\psi_n (x)$ is the corresponding eigenstate at $x$. In the case $T \gg \hbar / \Delta E$, where $\Delta E$ is the energy gap between the ground state and the first excited state, we have 
\begin{equation}
\left|\psi_0 \left( x\right) \right|^2 \propto K_E \left(x, T; x, 0 \right).
\end{equation}
Therefore, the normalized Euclidean propagator leads to the ground-state probability density $\left|\psi_0 \left( x\right) \right|^2$.

\subsection{Harmonic oscillator}
First in Fig.~\ref{fig:HO_Gen} we present the results for the prototypical harmonic oscillator.
Figure~\ref{fig:HO_Gen}(a) shows the estimated Euclidean propagator (or the kernel) after the trace normalization (blue dots), with fixed $x_i = 0$. The propagator trace is defined as 
\begin{equation}
\mathrm{tr} K_E = \int \mathrm{d}x \; K_E (x, T; x, 0),
\end{equation}
which is estimated through integrating over the smooth fit of VFPG results for the diagonal propagator. 
Figure~\ref{fig:HO_Gen}(b) plots the ground-state probability density obtained from the diagonal propagator (blue dots). The blue dashed line is a Gaussian fit of the VFPG observations. In Fig.~\ref{fig:HO_Gen}(a, b), the VFPG observations exhibit good agreement with the (red dashed) analytic exact results, which are obtained via Wick rotating the real-time 
propagator~\cite{RichardArxiv2000}. 

Note that we use a stochastic gradient descent algorithm to train the network (with Adam optimizer and learning rate $\eta = 1 \times 10^{-4}$), with $2048$ data points sampled from the $2\mathrm{D}$ latent space and a batch size of $128$. Such a training algorithm causes statistical noises in the final result of ${\cal{F}}_{\phi}$. Following the error analysis formula in Ref.~\cite{NicoliPRL2021} and with Eq.~(\ref{eq:est_prop}), the error bars (two-standard-deviations) in Fig.~\ref{fig:HO_Gen}(a, b) are estimated from the training noises (standard deviation $\delta {\cal{F}}_{\phi}$) of the free energy, with $\delta K_E = K_E \ \delta {\cal{F}}_{\phi} / \sqrt{N_r}$, in $N_r = 10$ independent runs of the training with each up to $3000$ epochs.

For the purpose of an intuitive understanding, we plot in Fig.~\ref{fig:HO_Gen}(c) an example of the distribution of the generated FPs in the plane of the variational log-probability vs the action. The approximate paths are distributed along the red solid line, which is the training target of the VFPG (In the case that ${\cal{F}}_{\phi}$ is unbiased for $-\hbar \mathrm{ln}Z$), i.e., the exponential distribution (\ref{eq:PDF_Euclidian_FP}). The red solid line is also the location of FPs that exactly satisfy $F_{\phi} \left[x(\tau)\right] = {\cal{F}}_{\phi}$, and the color of the images of the generated FPs (from blue to red) encodes the distance $d = \left| F_{\phi} \left[x(\tau)\right] - {\cal{F}}_{\phi}\right|$ from the exact distribution. Note that both the two axes in Fig.~\ref{fig:HO_Gen}(c) are shifted for a universal target distribution, where $S_{E \mathrm{min}}$ is the smallest Euclidean action and $b = \left(- {\cal{F}}_{\phi} + S_{E \mathrm{min}} \right)/\hbar$. 
Figure~\ref{fig:HO_Gen}(d) shows an example of the behavior of the free energy (blue line) during the training process (only 1000 epochs are shown), where the width of the red shaded area is its two-standard-derivations in the generated path ensemble.

\begin{figure}[t]
\centerline{\includegraphics[height=5in,width=3in,clip]{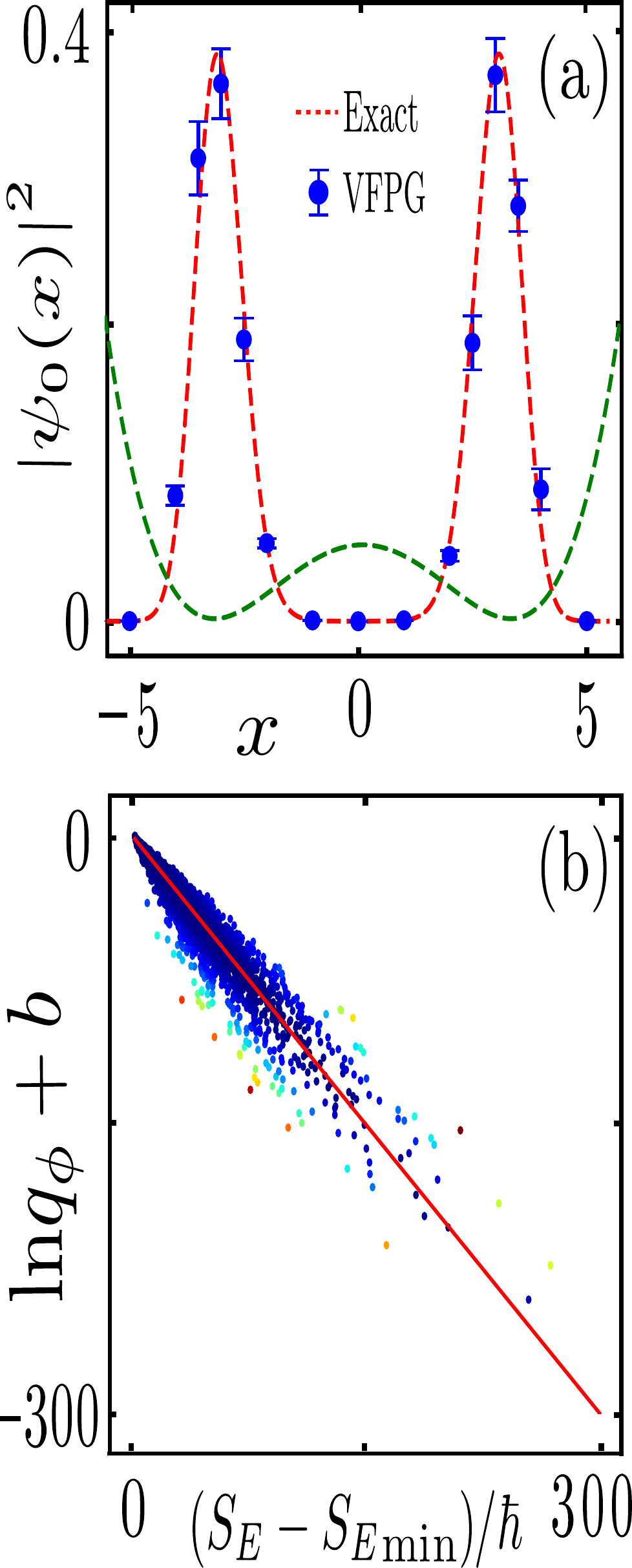}}
\caption{\textbf{Double-well potential $V(x) = \alpha x^4 + \beta x^2$.}
\textbf{(a)} Ground state wave function obtained from the Euclidean quantum propagator estimated via the variational Feynman path generator (VFPG), where the red dashed line is the result of the exact diagonalization. The green dashed line denotes the shifted and rescaled potential energy $\tilde{V}(x) = \left[ V(x) + 5 \right] / 100$.
\textbf{(b)} An example of the distribution of $10^4$ generated Feynman paths (with $x_i = x_f = x = 3$) in the plane of the log-probability and the action $S_E$, where both axes are shifted, with $S_{E \mathrm{min}}$ the minimum action and $b = \left(- {\cal{F}}_{\phi} + S_{E \mathrm{min}} \right)/\hbar$. The red solid line represents the training target (\ref{eq:PDF_Euclidian_FP}). It is also the location of Feynman paths that exactly satisfy $F_{\phi} \left[x(\tau)\right] = {\cal{F}}_{\phi}$, and the color (from blue to red) indicates the distance $d = \left| F_{\phi} \left[x(\tau)\right] - {\cal{F}}_{\phi}\right|$ from the exact distribution. Representative parameters used are: $\alpha = 0.05$, $\beta = -1$, $N_{\tau} = 32$, $\hbar = 1$, $m = 1$ and $T = 2$, respectively.
}
\label{fig:AHO_Gen}
\end{figure}

\subsection{Double-well potential} 
In Fig.~\ref{fig:AHO_Gen} we show the results from the VFPG for a double-well potential $V(x) = \alpha x^4 + \beta x^2$, with $\alpha = 0.05$, and $\beta = -1$. Figure~\ref{fig:AHO_Gen}(a) plots the estimated ground-state probability densities (blue dots), which again exhibit good agreement with the result from the exact diagonalization (red dashed). It has two peaks centered at the two respective minima of the double-well potential. Details for training and the error estimation formula are the same as those stated above for the harmonic oscillator. Again, the error bars (two-standard-deviations) here are estimated from the training noises (standard deviation $\delta {\cal{F}}_{\phi}$) of the free energy, in $N_r = 10$ independent runs of the training with each up to $3000$ epochs, as in Fig.~\ref{fig:HO_Gen}(a, b).
 
Shown in Fig.~\ref{fig:AHO_Gen}(b) is an example of the cluster of generated FPs in the plane of the variational log-probability vs the action, where the representative position value is taken as $x = 3$. As in Fig.~\ref{fig:HO_Gen}(c), the generated paths are distributed around the training target of the VFPG (red solid line). Details and other physical parameters can be found in the caption of Fig.~\ref{fig:AHO_Gen}.

\section{Discussion and outlook} 
The propagator plays the role of the kernel for the quantum evolution. By virtue of the FP integral representation, here the kernel as well as the ground-state density can be efficiently estimated by generating FPs via parallel sampling. Instead of the generative modeling of quantum fields~\cite{PhysRevLett.121.260601,AlbergoPRD2019,PhysRevResearch.2.023369,PhysRevLett.125.121601,NicoliPRL2021,PhysRevD.104.114507,MatijaArxiv2020,arxiv:2201.08862v2}, here we focus on generating Feynman paths for quantum mechanical systems (i.e., trajectories or sequences as in Refs.~\cite{GFlow_1,GFlow_2}), with a fixed starting and terminating position. There are two advantages of the proposed VFPG: First, the sample complexity in the $2\mathrm{D}$ latent space is lower compared with normalizing-flow models~\cite{NICE2014,RealNVP2016,AlbergoPRD2019,PhysRevLett.125.121601,PhysRevD.104.114507,NicoliPRL2021,MatijaArxiv2020,arxiv:2201.08862v2}, which are bijections between the latent space and the path space, and require a much higher latent dimension (same as the output dimension); Second, the sampling of the output paths is performed in a parallel manner and therefore is more efficient than Markov chain Monte Carlo (MCMC) 
methods~\cite{InferringDM2019}, which iteratively generate path samples from the exact target path distribution in Eq.~(\ref{eq:PDF_Euclidian_FP}). Moreover, the Monte Carlo method cannot give a direct estimation of the partition function (i.e., the propagator) as in our work.

In the limit $\hbar \,\rightarrow \, 0$, the loss function in~(\ref{eq:Generator_Loss}) will be dominated by the action term, and we find that Feynman paths generated by our RNN decoder (the VFPG) collapse to the minimal-action path, which can give a different perspective to understand the quantum-classical correspondence. In contrast, normalizing flows are bijections, which cannot be used for this purpose. As a matter of fact, we also produced ourselves some results on estimating the propagator of a harmonic oscillator with flow-based models (e.g., the real-valued non-volume
preserving (real NVP) transformations in Ref.~\cite{RealNVP2016}), but so far the performance is not as good as the results presented in this paper (therefore those are not shown here). More elaborations in this respect and detailed comparisons between the current VFPG and normalizing flows are left for future work. 

In addition, the variance reduction for discrete variable systems as in Ref.~\cite{WuPRL2019}, inspired from reinforcement learning, does not apply well to the continuous variables considered here. An alternative systematic variance reduction is required, which is particularly important when the numerical value of the final free energy is small. Other possible generalizations may include extending the current results to higher spatial dimensions and to more complex systems, as well as investigating the generative modeling of FPs in the Lorentzian space-time.

\section{Summary}
We delivered the concept of FP generators for modeling Euclidean quantum mechanical FPs with \emph{fixed endpoints} from the latent space. A variational realization with the recurrent neural network is performed, and as a proof-of-principle demonstration, the quantum propagators (or kernel functions) are efficiently estimated for both a harmonic oscillator and an anharmonic potential. Our work paves the way toward deep generative modeling of FPs with fixed starting and terminating points, respectively, and can provide a future fresh perspective to understand the quantum-classical correspondence through deep learning.

All the data and the code (in PYTHON and TENSORFLOW) for generating the results in this work are available upon request to the authors.

\emph{Note added.}
Recently, we noted the generative flow network (GFlowNet)~\cite{GFlow_1,GFlow_2}, which generates a set of paths from a starting state to a terminating one on a graph, but with a different learning objective (compared to the path generator in our work) and for different tasks (in particular, molecule generation).

\begin{acknowledgments}
The computation was performed in the RIKEN supercomputer (HOKUSAI) system. 
We acknowledge Tao Liu, Zheng-Yang Zhou and Yu-Ran Zhang for helpful discussions. 
We thank Enrico Rinaldi for critical reading of the manuscript.
F.N. is supported in part by: 
Nippon Telegraph and Telephone Corporation (NTT) Research, 
the Japan Science and Technology Agency (JST) [via the Quantum Leap Flagship Program (Q-LEAP), 
the Moonshot R\&D Grant Number JPMJMS2061, 
the Japan Society for the Promotion of Science (JSPS) 
[via the Grants-in-Aid for Scientific Research (KAKENHI) Grant No. JP20H00134], 
the Army Research Office (ARO) (Grant No. W911NF-18-1-0358), 
the Asian Office of Aerospace Research and Development (AOARD) (via Grant No. FA2386-20-1-4069), and the Foundational Questions Institute Fund (FQXi) via Grant No. FQXi-IAF19-06.
\end{acknowledgments}

\appendix

\section{modeling the density of paths with variational recurrent neural networks}
\label{sec:A}

Here the Feynman path $x(\tau)$ is approximately represented by a discrete $N_{\tau}$-dimensional vector of a time series $\{ x \left( \tau_k \right) \}^{N_{\tau} - 1 }_{k=0}$, with $\tau_k = k \times \delta \tau$, where $\delta \tau$ is a fixed time step. So the time interval $\tau \in [0, T]$ is sliced into $N_{\tau}$ discrete points with $ T = \left( N_{\tau} - 1 \right) \delta \tau $, while the spatial dimensions are continuous, i.e., not discretized. This is a time-domain lattice approximation of the Feynman path, which is suitable for machine learning models. It can be a good approximation when a small time step $\delta \tau$ is used, and the dimension $N_{\tau}$ of the time series is proportional to the total time $T$.

Here we model the output density $q_{\phi} \left[ x(\tau) \right]$ of Feynman paths with a recurrent neural network (see Fig. 2 in the main text), where $\phi$ represents a set of neural-network parameters. In such a model, the probability density of a discrete path $\{ x \left( \tau_k \right) \}^{N_{\tau} - 1 }_{k=0}$ is given by an autoregressive form~\cite{GoodfellowDL2016}
\begin{eqnarray}
q_{\phi} \left[ x(\tau) \right] = \prod_{k=0}^{N_{\tau} - 1} q_k \left[ x \left( \tau_k \right) | x \left( \tau_{<k} \right) \right],
\end{eqnarray}
where the factor distribution $q_k \left[ x \left( \tau_k \right) | x \left( \tau_{<k} \right) \right]$ denotes the probability density of positions at $\tau_k$ conditioned on its previous times stamps, with ${<k}$ denoting $i = 0, 1, ..., k-1$, and the first and last positions are constrained to match the specified endpoints, $x_i$ and $x_f$, respectively. This factor distribution at each time stamp is parametrized by a Gaussian mixture model (GMM), which is a universal approximator of densities~\cite{GoodfellowDL2016}. In the GMM, we have 
\begin{equation}
q_k \left[ x \left( \tau_k \right) | x \left( \tau_{<k} \right) \right] 
= \sum_{j=1}^{N_c} \gamma_{j}^{(k)} \ {\cal{N}} \left( x \left( \tau_k \right); \mu_{j}^{(k)}, \sigma_{j}^{(k)}\right),
\end{equation}
where the mixing weight $\gamma_{j}^{(k)}$, the mean $\mu_{j}^{(k)}$ and the standard deviation 
$\sigma_{j}^{(k)}$ of the component Gaussian are the output of the recurrent neural network at the 
$k$-th unit [see Fig.2(b) in the main text], and therefore they intrinsically have the conditional dependence on the values in the previous time stamps. The number of Gaussian components is $N_c$ 
(we set $N_c$ to be the batch size in our code). The mixing weight satisfies $0 \le \gamma_{j}^{(k)} \le 1$ and $\sum_{j=1}^{N_c} \gamma_{j}^{(k)} = 1$, which is realized by a softmax function. 
Compared to the MCMC sampling, here the generation of Feynman paths can be obtained by combinations of parallel samplings at each time stamps.

\section{Total loss function.} 
\label{sec:B}

The training of the Feynman path generator involves the minimization of the KL divergence between the variational density $q_{\phi} \left[ x \left( \tau \right) \right]$ of paths and the target distribution $P_E \left[ x \left( \tau \right) \right]$, which equals to the (shifted) variational free energy in the main text. The constraints on the two endpoints are added to the KL loss as penalties. In the GMM formulation of $q_{\phi} \left[ x \left( \tau \right) \right]$, the total loss can be written as 
\begin{equation}
 {\cal{L}}_{\mathrm{total}} = {\cal{L}}_{\mathrm{KL}} + {\cal{L}}_1 + {\cal{L}}_2 ,
\end{equation}
where
\begin{eqnarray}
{\cal{L}}_{\mathrm{KL}} &=& \mathop{\mathbb{E}}_{x \sim q_{\phi}} F_{\phi}[x(\tau)] / \hbar,
\end{eqnarray}
with $F_{\phi}[x(\tau)] = S_E \left[x(\tau)\right] + \hbar \log q_{\phi}[x(\tau)]$, and
\begin{eqnarray}
{\cal{L}}_1 &=& N_{\tau}\sum_{j=1}^{N_c} \left[ \mu_{j}^{(k=0)} - x_i \right]^2 + \left[ \sigma_{j}^{(k=0)} \right]^2 \\ 
\nonumber
{\cal{L}}_2 &=& N_{\tau}\sum_{j=1}^{N_c} \left[ \mu_{j}^{(k=N_{\tau} - 1)} - x_f \right]^2 + \left[\sigma_{j}^{(k=N_{\tau} - 1)} \right]^2,
\end{eqnarray}
where $x_i$ and $x_f$ are the initial and final positions of all possible paths, respectively.


\begin{thebibliography}{68}
\bibitem [{sta()}]{stationary_note}%
  \BibitemOpen
  \href@noop {} {}\bibinfo {note} {For the Euclidean case, the most probable
  path should also satisfy $\delta ^2 S > 0$.}\BibitemShut {Stop}%
\end{thebibliography}
%

\end{document}